# Verifying the Correctness of Analytic Query Results

Masoud Nosrati and Ying Cai

**Abstract**— Data outsourcing is a cost-effective solution for data owners to tackle issues such as large volumes of data, huge number of users, and intensive computation needed for data analysis. They can simply upload their databases to a cloud and let it perform all management work, including query processing. One problem with this service model is how query issuers can verify the query results they receive are indeed correct. This concern is legitimate because, as a third party, clouds may not be fully trustworthy, and as a large data center, clouds are ideal targets for hackers. There has been significant work on query result verification, but most consider only simple queries where query results can be attained by checking the raw data against the query conditions directly. In this paper, we consider the problem of enabling users to verify the correctness of the results of analytic queries. Unlike simple queries, analytic queries involve ranking functions to score a database, which makes it difficult to build data structures for verification purposes. We propose two approaches, namely *one-signature* and *multi-signature*, and show that they work well on three representative types of analytic queries, including *top-k*, *range*, and *KNN* queries, through both analysis and experiments.

**Index Terms**— analytic queries, query result verification, signature mesh, I-tree

---------------------------◆---------------------------

## 1 INTRODUCTION

ANALYTIC queries involve utility functions in data retrieval. A *top-k* query, for example, applies a function to compute a score for each record in a database and returns those whose scores are among the $k$ highest [1]. Another example is *range* query, i.e., finding the records whose scores under a utility function are within a specific range. Such queries are essential to applications such as finding patients with a high risk of breast cancer [2], diabetes and/or Alzheimer [3], [4], customers with minimal financial risk [5], and potential money laundering crimes [6], where functions are used to score a database. In addition to the ones mentioned above, various types of analytic queries have been developed so far, including *reverse top-k* [7], *scalar product* [8], *maximum rank* [9], and *global immutable region* [10], just to name a few. Despite their differences, analytic queries are common in involving one or more utility functions in query processing.

In this paper, we consider the problem of verifying the correctness of analytic query results. Specifically, we consider a data outsourcing model that consists of three parties: data owner, cloud service provider, and data users. The data owner uploads its database to a cloud server, and the data users send their queries to the server. Upon receiving a query, the server processes the query and returns the result to the query issuer. While delegating the responsibility of data management to the server, the data owner wants to provide a mechanism for data users to verify whether or not the query results they receive are indeed correct. Here a query result is said to be correct if it is *sound* and *complete*. The former means every record included in the query result is original in the database and satisfies the query condition, whereas the latter means all records in the original database that satisfy the query condition are included in the query result. The concern that the server may return incorrect query results is legitimate. There may be inside attacks, where the server is intentionally configured to return incomplete query results for reasons such as saving costs. There may also be external attacks. Large data centers like clouds are ideal targets for hackers. Moreover, the networks which connect the server and data users are also subject to attacks.

The problem of query result verification was first studied by Devanbu et al. [11] in the context of simple range query, where users retrieve data whose value is within a specific range. In their approach, a data owner sorts the data items to be outsourced, applies a one-way hash function on the sorted values, and then builds a *Merkle Hash tree* (MH-tree) [12] on the hash results, where the root of the tree is signed with the data owner's private key. The data items and the MH-tree are then uploaded to the cloud server. When processing a range query, the server returns not only the query result but also a *verification object* (VO), a piece of the MH-tree containing the signed root that can serve as the proof that the data items in the query result are indeed sound and complete. An alternative approach was later proposed by Pang et al. [13], which builds a *signature chain* on the one-way hash values of the sorted data items. The two verification data structures, namely MH-tree and signature chain, have since been extended to support other types of queries such as multi-dimensional range query and spatial queries [14], [15], [16], [17], [18], [19].

---

● *Masoud Nosrati and Ying Cai are with the Department of Computer Science, Iowa State University, Ames, IA 50011.*
*E-mail: {nosrati, yingcai}@iastate.edu*

Most queries considered in existing research are simple queries in the sense that they can be processed by checking the raw data directly against a query condition. Given a range query, for example, one can just check each data item and return those that are in the specified range. Analytic queries, however, are complex as they involve utility function. A *top-k* query, for example, returns the $k$ data items with the highest scores under a given utility function. The utility function, which is supplied by the query issuer, is not known when the raw data is uploaded to the server. Since the data owner cannot precompute the score for each data item, it becomes a great challenge to build verification data structures. To our knowledge, [20] is the only work that considers the queries with utility functions. Their proposed scheme converts each data item in the database into a continuous function and partitions the function domain into a number of subdomains. Since in each of these subdomains the functions can be sorted based on their output, it becomes possible to build a *signature mesh* on the sorted lists for verification. This work, however, has some performance issues. To process a query, the server performs a linear search on the subdomains, the number of which can be very large. For a database of $n$ records ranked by $d$-variable linear functions, the number of subdomains can be up to $O(n^{2d})$. This number is even higher when more complex ranking functions are used. While the efficiency in constructing verification objects on the server side is of great concern, the constructed verification objects may also consist of a large number of signatures and incur high computation cost in verification. This issue is a concern to data users, especially when mobile devices are used. In this paper, we address these problems and make the following contributions:

- We propose a generic solution for verifying the correctness of analytic query results. While our discussion limits itself to three well-known query types, including *top-k*, *range*, and *KNN* queries, for the purpose of comparing with existing signature mesh head to head, the proposed technique can be used to support other types of queries.

- Our technique is more practical. We extend and integrate the concepts of *Intersection-tree* (I-tree) and Merkle Hash-tree (MH-tree) to index the database for efficient query processing and construction of verification objects. The proposed technique also minimizes the cost for data users to perform query result verification. Detailed algorithms are given to illustrate how to build this verification data structure, construct verification objects, and verify the query results.

- The performance of the proposed techniques is studied through both analysis and experiments. We prove that the proposed technique can indeed allow users to perform query result verification. As for the costs such as constructing verification objects and verifying query results, our extensive experiments indicate that our approach outperforms the existing approach to a large extent.

The rest of this paper is organized as follows. In section 2, we give the background of this research, including system model, adversary model and security goal, and related

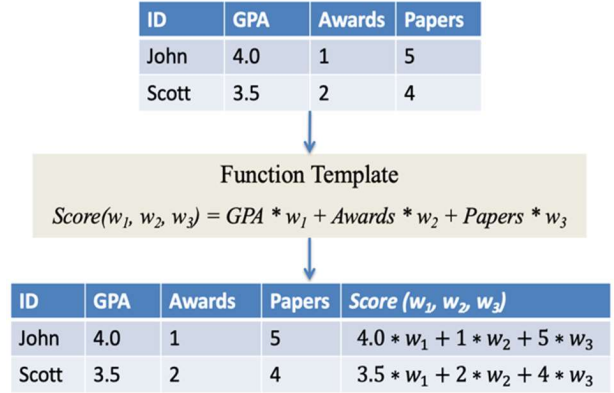

| ID | GPA | Awards | Papers |
|----|-----|--------|--------|
| John | 4.0 | 1 | 5 |
| Scott | 3.5 | 2 | 4 |

Function Template

$Score(w_1, w_2, w_3) = GPA * w_1 + Awards * w_2 + Papers * w_3$

| ID | GPA | Awards | Papers | $Score\ (w_1, w_2, w_3)$ |
|----|-----|--------|--------|--------|
| John | 4.0 | 1 | 5 | $4.0 * w_1 + 1 * w_2 + 5 * w_3$ |
| Scott | 3.5 | 2 | 4 | $3.5 * w_1 + 2 * w_2 + 4 * w_3$ |

Fig. 1. Each record is interpreted as a math function.

work in query result verification and I-tree. Our proposed technique is presented in Section 3, and the performance studied in Section 4. We conclude this paper in Section 5.

## 2 BACKGROUND

### 2.1 System Model

Consider a data owner who outsources a database to a cloud server and allows data users to perform analytic queries over the database. For this purpose, in addition to the database, the owner also uploads a template of utility functions to be used in queries. With the template, the server interprets each record as a math function.

To illustrate, consider the table showed in Fig. 1, where each tuple records an applicant's ID, GPA, the number of awards, and the number of papers. The template of utility functions for this table is $Score(w_1, w_2) = GPA \times w_1 + Award \times w_2 + Paper \times w_3$. Accordingly, the server interprets each tuple $r_i$ as a math function $Score_i(w_1, w_2, w_3) = GPA_i \times w_1 + Award_i \times w_2 + Paper_i \times w_3$, where $w_1$, $w_2$, and $w_3$ are variables, and $GPA_i$, $Award_i$, and $Paper_i$ are corresponding attribute values of the record $r_i$.

Since this paper concerns only analytic queries, we will simply say the outsourced database is a set of math functions $\{f_1, f_2, ..., f_n\}$, where $f_i$ is the function corresponding to record $r_i$ in the database under the given utility function template. When causing no ambiguity, we will also use the terms record and function interchangeably. The functions have the same set of variables $X = (x_1, x_2, ..., x_d)$. For example, the functions in the applicant table have three variables, i.e., $X = (w_1, w_2, w_3)$.

Data users send their queries to the server. The server processes the queries and returns the results to users. In this paper, we consider three types of analytic queries, which are representative and popularly used:

- A *top-k query* $q = (X, k)$ retrieves all records $r_i$ whose function output $f_i(X)$ is among the top $k$.

- A *range Query* $q = (X, l, u)$ retrieves all records $r_i$ such that $l \le f_i(X) \le u$. We will refer to $l$ and $u$ as the query's lower and upper boundaries, respectively.

- A *KNN query* $q = (X, k, y)$ retrieves all records $r_i$ whose function output $f_i(X)$ is among the $k$ neighbors that nearest to the given value $y$.

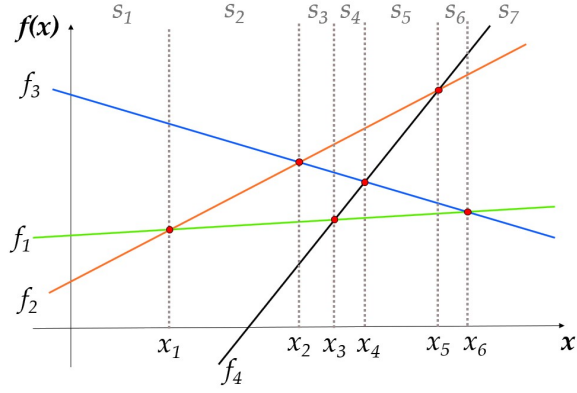

(a) Intersections of 4 linear functions

| (−∞,x₁) | [x₁,x₂] | [x₂,x₃] | [x₃,x₄] | [x₄,x₅] | [x₅,x₆] | [x₆,+∞) |
|---|---|---|---|---|---|---|
| max | max | max | max | max | max | max |
| $f_3$ | $f_3$ | $f_2$ | $f_2$ | $f_2$ | $f_4$ | $f_4$ |
| $f_1$ | $f_2$ | $f_3$ | $f_3$ | $f_4$ | $f_2$ | $f_2$ |
| $f_2$ | $f_1$ | $f_1$ | $f_4$ | $f_3$ | $f_3$ | $f_1$ |
| $f_4$ | $f_4$ | $f_4$ | $f_1$ | $f_1$ | $f_1$ | $f_3$ |
| min | min | min | min | min | min | min |

(b) Sorted functions lists for each subdomain

| (−∞,x₁) | [x₁,x₂] | [x₂,x₃] | [x₃,x₄] | [x₄,x₅] | [x₅,x₆] | [x₆,+∞) |
|---|---|---|---|---|---|---|
| $\mathrm{Sig}(\max\mid r_3)$ | $\mathrm{Sig}(\max\mid r_2)$ | | | | $\mathrm{Sig}(\max\mid r_4)$ | |
| $\mathrm{Sig}(r_3\mid r_1)$ | $\mathrm{Sig}(r_3\mid r_2)$ | $\mathrm{Sig}(r_2\mid r_3)$ | | $\mathrm{Sig}(r_2\mid r_4)$ | $\mathrm{Sig}(r_4\mid r_2)$ | |
| $\mathrm{Sig}(r_1\mid r_2)$ | $\mathrm{Sig}(r_2\mid r_1)$ | $\mathrm{Sig}(r_3\mid r_1)$ | $\mathrm{Sig}(r_3\mid r_4)$ | $\mathrm{Sig}(r_4\mid r_3)$ | $\mathrm{Sig}(r_2\mid r_3)$ | $\mathrm{Sig}(r_2\mid r_1)$ |
| $\mathrm{Sig}(r_2\mid r_4)$ | $\mathrm{Sig}(r_1\mid r_4)$ | | $\mathrm{Sig}(r_4\mid r_1)$ | $\mathrm{Sig}(r_3\mid r_1)$ | | $\mathrm{Sig}(r_1\mid r_3)$ |
| $\mathrm{Sig}(r_4\mid \min)$ | | | | $\mathrm{Sig}(r_1\mid \min)$ | | $\mathrm{Sig}(r_3\mid \min)$ |

(c) Signature mesh

Fig. 2. An example of signature mesh built for four functions

## 2.2 Adversary Model and Security Goal

In response to a query, the server processes and returns the result to the query issuer. The **adversary model** is as follows. The server may be configured to send out a wrong query result, intentionally or unintentionally, by an insider or a malicious intruder. Or the server sends out the correct result, but then it is modified during the network transmission. In short, we simply assume the query result received by a data user may be incorrect for whatever causes. Our **security goal** is to provide a mechanism for users to verify that the query results they receive are indeed correct. A query result $R(q)$ is said to be correct if it satisfies two requirements:

- **Soundness**: Every data item in $R(q)$ appears in the original database and satisfies the query condition.
- **Completeness**: Every data item in the original database that satisfies the query condition is included in $R(q)$.

## 2.3 Related work

### 2.3.1 Query Result Verification

There has been significant work on developing hardware (e.g., Trusted Platform Module (TPM) [21], Hardware Security Module (HSM) [22], [23]) to support secure computation. Our research is interested in software-based solutions, which do not require special chips. Existing approaches, such as MH-tree [11] and Signature Chain [13], are software-based, but as mentioned, they were developed for simple queries and cannot be applied directly for verifying the results of analytic queries. The utility functions used in analytic queries are not known when the raw data is uploaded to the server. Since the data owner cannot precompute the score for each data item, verifying query results becomes significantly challenging.

The above problem was first studied in [20], where a *signature mesh* approach was developed. The solution is based on the theorem of *function sortability*: Given a set of functions, their domain $D$ can be partitioned into a set of disjoint subdomains $D = S_1 \cup S_2 \cup \ ... \cup S_n$ such that for each subdomain $s_i (1 \leq i \leq n)$, the functions have the same order when sorted according to their scores for all input $X$ in $s_i$.

To illustrate, consider four univariate linear functions, $f_1(x)$, $f_2(x)$, $f_3(x)$, and $f_4(x)$, shown in Fig. 2a. The four lines intersect on six points, including $x_1$, $x_2$, $x_3$, $x_4$, $x_5$, and $x_6$, which partition their domain $S$ into seven subdomains, $(-\infty, x_1)$, $[x_1, x_2]$, $[x_2, x_3]$, $[x_3, x_4]$, $[x_4, x_5]$, $[x_5, x_6]$, and $[x_6, +\infty)$. It is clear that in each of these subdomains, the functions can be sorted based on their output. For example, for all $x$ in $[x_1, x_2]$, we have $f_3(x) \geq f_2(x) \geq f_1(x) \geq f_4(x)$. That is, the order of the four functions is the same for all $x$ in the range $[x_1, x_2]$.

In general, a set of $n$ linear functions with $d$ variables can have up to $n \times (n - 1)$ intersections, which together partition their domain into $O(n^{2d})$ subdomains, each being a hyperplane in a $d$-dimension domain space. The intersections of nonlinear functions are more complicated, but regardless of their complexity, it remains true that the intersections of a set of functions partition their domain into a number of subdomains. In each of these subdomains, the functions can be sorted according to their output.

In light of the theorem of function sortability, the idea of signature mesh becomes straightforward. Given a set of functions, the data owner performs the following steps: 1) Compute the intersections of the functions and the subdomains created by these intersections; 2) Sort the functions for each subdomain; 3) Add two special tokens (*min* and *max*) to each function list to indicate that they are the first and the last records (illustrated in Fig. 2b); 4) Create a signature chain for each function list. Let $f_1(\cdot) \leq f_2(\cdot) \leq \ldots \leq f_n(\cdot)$ be the sorted list for a subdomain $s_i$. For each pair of consecutive functions $f_j(\cdot)$ and $f_{j+1}(\cdot)$ in the list, the data owner computes a digest $H(H(r_j)|H(r_{j+1})|B_i)$, where $H(\cdot)$ is a one-way hash function, $r_j$ and $r_{j+1}$ are the records corresponding to $f_j$ and $f_{j+1}$, respectively, $B_i$ is the set of intersections that defines subdomain $s_i$. The digest is signed with the data owner's private key, which produces signature $Sig_{<s_i>}(r_j|r_{j+1}) = Sig\left(H(H(r_j)|H(r_{j+1})|B_i)\right)$.

Note that two functions that are consecutive in one subdomain may remain consecutive in a number of consecutive subdomains. In this case, only one signature is needed for the pair of functions across all these subdomains. This reduces the number of signatures and turns the whole set of signatures into a signature mesh, an example of which is showed in Fig. 2c.

Given two records $r_j$ and $r_{j+1}$, subdomain $s_i$, and $Sig_{<s_j>}(r_j|r_{j+1})$, a user can recompute the digest $H(H(r_j)|H(r_{j+1})|B_i)$ and decrypt the $Sig_{<s_i>}(r_j|r_{j+1})$ with the data owner's public key. If the two pieces of data match, the user can be convinced that: 1) the two records are original in the database, and 2) they are consecutive in the function list sorted for the subdomain $s_i$. As such, the signature mesh can be used for query result verification. Consider a range query $q = (X, l, u)$. The server first finds the subdomain that contains the user input $X$ and the corresponding sorted list for the function. On the list, the server finds the sub-list of the functions whose output under $X$ is in between $l$ and $u$. Let $f_a(\cdot) \leq f_{a+1}(\cdot) \leq \ldots \leq f_{b-1}(\cdot) \leq f_b(\cdot)$ be the sublist. Then the server sends the query result $R(q) = \{r_a, r_{a+1}, \ldots, r_{b-1}, r_b\}$, where $r_i$ is the record corresponding to $f_i$ and the corresponding verification object, which includes 1) $r_{a-1}$ and $r_{b+1}$, two additional records that are immediate left and right to the sublist, 2) the signatures for every pair of the consecutive functions in the list, and 3) the subdomain that contains $X$.

### 2.3.2 Intersection Tree (I-tree)

The intersection of two functions $f_i(X)$ and $f_j(X)$ is a hyperspace $\{X \mid f_i(X) - f_j(X) = 0\}$, which partitions their domain into two subdomains, say *above* and *below*. As such, the intersections of a set of functions partition their domain into a set of subdomains. The I-tree was designed to compute and index these subdomains [14].

An internal node in an I-tree, called an *intersection node*, has a form of $(f_i, f_j, a, b)$. It records the fact that two functions $f_i$ and $f_j$ intersect in some domain **X**. That is, there exists at least one $X$ in **X** such that $f_i(X) - f_j(X) = 0$. There is no need to know the boundary of **X**, except for the root node, whose **X** is the entire domain specified by the data owner. The intersection of $f_i$ and $f_j$, denoted $I_{i,j}$, partitions **X** into two subdomains *above* and *below*, which are represented by two pointers $a$ and $b$, respectively. The subdomain *above* consists of all inputs $X$ such that $f_i(X) - f_j(X) \geq 0$, whereas *below*, all inputs $X$ such that $f_i(X) - f_j(X) < 0$. If *above* is further partitioned by another intersection $I_{p,q}$, then $a$ links to the intersection node representing $I_{p,q}$. Otherwise, *above* is a subdomain where the functions can be strictly sorted according to their output. In this case, $a$ links to a *subdomain node*. A subdomain node is a tuple of $(s_i, L_i)$, where $s_i$ refers to a subdomain and $L_i$ the list of functions sorted for $s_i$. The same rule applies to the subdomain *below*: If it is further partitioned, it links to another intersection node; otherwise, it links to a subdomain node. All subdomain nodes are leaf nodes.

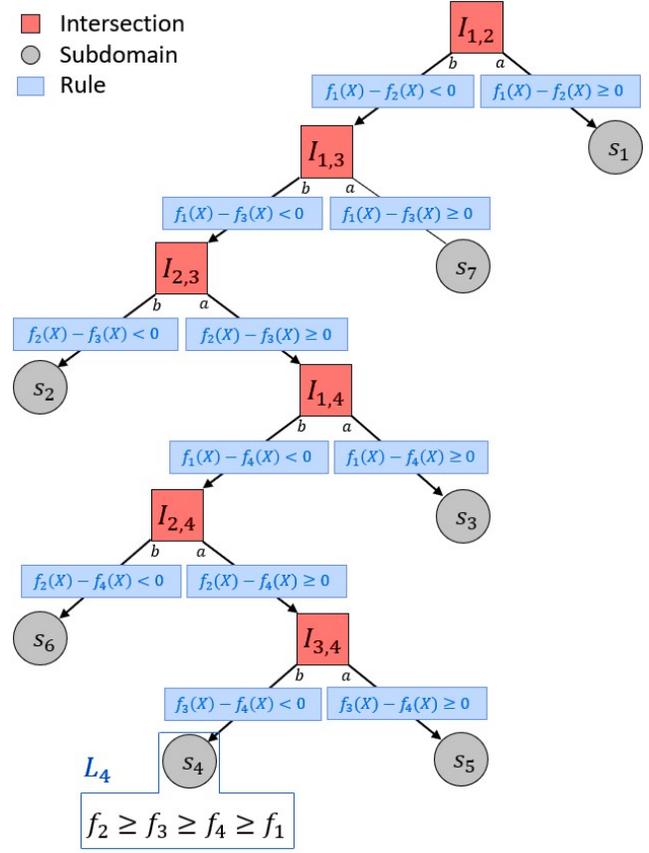

Fig. 3. An example of I-tree (for the functions in Fig. 2a)

Fig. 3 shows an I-tree built for the four functions illustrated in Fig. 2a. The root node records $I_{1,2}$, the intersection of $f_1$ and $f_2$. The two subdomains created by this intersection are represented by two subtrees linked by $a$ and $b$, respectively. The subdomain that is above $I_{1,2}$ is then partitioned by another intersection $I_{1,3}$, while the subdomain that is below $I_{1,2}$ is not partitioned further. This information is recorded by two nodes at layer 2, i.e., node with $I_{1,3}$ and node with $s_1$. The subdomain represented by $I_{1,3}$ is further partitioned by the intersection $I_{2,3}$ into two subdomains, represented by two nodes linked by $a$ and $b$, respectively.

Once the I-tree is built, we can then sort the functions and have a pointer from each subdomain node linking to its corresponding sorted function list. For simplicity, Fig. 3 shows only one subdomain node (i.e., $s_4$) that links its sorted function list.

I-tree supports efficient search of subdomains. Given a function input $X$, the search starts from the tree root. It first checks which half-space $X$ belongs to. If $X$ belongs to *above*, then follow the link $a$; otherwise, follow the link $b$. This process is repeated until a subdomain node is reached, which has a link to a sorted function list. This feature allows the I-tree to support highly efficient processing of analytic queries, which will be explained in detail later.

## 3 PROPOSED SOLUTIONS

To our knowledge, the signature mesh approach is the only approach for verifying the results of analytic queries. This scheme, however, incurs significant run-time overheads to both server and data users. Consider a top-k query $q = (X, k)$, which retrieves all records whose scores under input $X$ is among the top $k$. To construct a verification object, the server needs to perform a linear search to find the subdomain that contains $X$. As mentioned, for a database of $n$ records ranked by $d$-variable linear functions, the number of subdomains is $O(n^{2d})$. This number is even higher when more complex ranking functions are used. As such, the cost of constructing a verification object for a top-k query (and other queries such as range and KNN) can be prohibitively high. Moreover, when a verification object consists of a large number of signatures, the cost of verification is also high, a great concern to battery-powered mobile devices.

To address these problems, we propose a new verification data structure called *Intersection and Function Merkle Hash tree* (IFMH-tree), which is an extension and combination of the concepts I-tree and MH-tree. It has two components. The first one, referred to as *Intersection Merkle Hash tree* (IMH-tree), supports efficient search and verification of the subdomain that contains a given function input. The second component, called *Function Merkle Hash-tree* (FMH-tree), supports efficient search and verification of the functions on a sorted function list. Depending on where to sign, the proposed technique has two versions, *one-signature* and *multi-signature*. In the following subsections, we discuss in detail how to build the IFMH-tree, construct verification objects, and verify the correctness of the query results.

### 3.1 Building IFMH-tree

To start with, we illustrate an IFMH-tree in Fig. 4 that is corresponding to the I-tree in Fig. 3. Note that for every node in the I-tree, we add a new attribute to store a hash value. And for every sorted function list, we build an FMH-tree and add a pointer in the corresponding subdomain node to link the tree. The steps of creating an IFMH-tree are as follows.

#### Step 1: Building an IMH-tree without hash values

Let $\{f_1, f_2, ..., f_n\}$ be the $n$ functions corresponding to the $n$ record in the database. There is a total of $n * (n - 1)$ pairs of functions $(f_i, f_j)$, where $1 \le i, j \le n$ and $i \ne j$. We create an empty I-tree root node with the domain given to the function variables. Then for every pair of the functions $(f_i, f_j)$, insert the intersection $I_{i,j}$ to the tree by creating an empty queue $Q$, adding the tree root to $Q$, and repeating

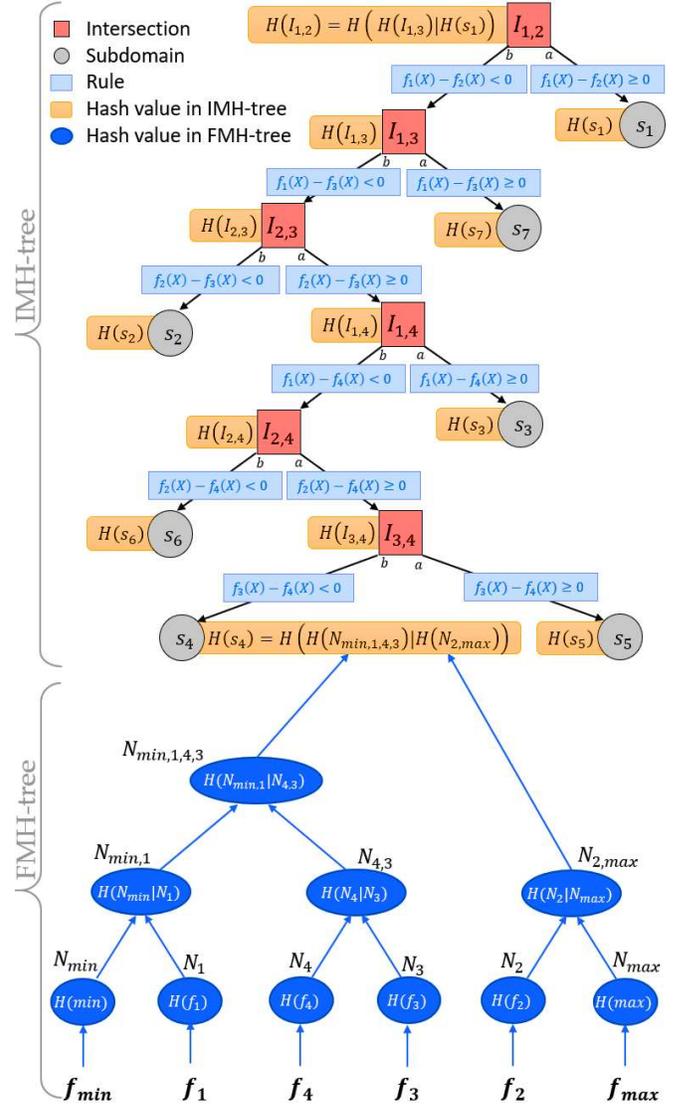

Fig. 4. An IFMH-tree Example

the following operations until $Q$ is empty:

- Dequeue a node from $Q$.  Let $N$ be this node, and **X** be the domain represented by $N$.
- Check if $I_{i,j}$ partitions **X.**  If this is true, handle two scenarios accordingly:
    - $N$ is a subdomain node: In this case, record $I_{i,j}$ in $N$, create two new subdomain nodes, and link them with $N$'s two pointers, $N.a$ and $N.b$, respectively.
    - N is an intersection node: In this case, enqueue $N$'s two child nodes, which are linked by $N$'s two pointers, $N.a$ and $N.b$, respectively.

After inserting all function pairs, then for each subdomain node $N$, sort the functions for the subdomain represented by $N$ and have $N$'s pointer link to the sorted function list.

Note in the above algorithm, whenever a new node is allocated, its hash value is set to be 0, as a default value that indicates the hash has not been computed yet; i.e., it is invalid.

***Step 2: Building an FMH-tree for each sorted function list***

In an FMH-tree, each node $N$ has four attributes, including $h$, $p$, $l$, and $r$, where $h$ stores a hash value, $p$, $l$, and $r$ are three pointers. Let $f_1(\cdot) \leq f_2(\cdot) \leq \ldots \leq f_n(\cdot)$ be a sorted function list. We add two special token functions, $f_{min}$ and $f_{max}$, to indicate the beginning and the end of this list, i.e., $f_{min}(\cdot) \leq f_1(\cdot) \leq f_2(\cdot) \leq \ldots \leq f_n(\cdot) \leq f_{max}(\cdot)$. For each function $f_i$, we compute a hash value $H(f_i)$, where $H(\cdot)$ is a one-way hash function, and create a tree node $N_i$, where $N_i.h = H(f_i)$, and three pointers are set to be null. We then build the tree layer by layer, starting from the bottom layer. Recall the nodes at the bottom layer are $N_1$, $N_2$, ..., $N_n$. We start from left to right and create a parent node for every two nodes on the list. That is, for $N_i$ and $N_{i+1}$, we create a new node $N_{i,i+1}$, and let $N_{i,i+1}.h = H(N_i.h \mid N_{i+1}.h)$, $N_{i,i+1}.l = N_i$, and $N_{i,i+1}.r = N_{i+1}$. For example, for both $N_{mi}$ and $N_1$ in Fig. 4, we set their $p$ pointers to their parent, i.e., $N_{min}.p = N_{min,1}$ and $N_1.p = N_{min,1}$.

We then move to $N_3$ and $N_4$, and so on so forth for all nodes in the list. If the number of nodes is odd, then the last node will be linked to the tree in the next round. This process is repeated until the new layer contains only one node. This node is the root of the FMH-tree. An FMH-tree for the function list linked by subdomain $S_4$ is showed in Fig. 4. Note that in the original MH-tree, the hash value stored at the root is signed with the data owner's private key. For now, we leave it unsigned.

***Step 3: Propagating the hash values to IMH-tree***

Recall that every subdomain node in I-tree has a pointer linking to its corresponding sorted function list. We add another pointer to each subdomain node to link it to the FMH-tree that was constructed for the function list. Without causing ambiguity, we will say every subdomain node has a hash value, which is equal to the hash value stored at the root of the FMH-tree linked by the subdomain node.

Also recall that every intersection node in an I-tree has two pointers, $a$ and $b$, linking two child nodes. We add a new attribute $h$ to store a hash value, which is initialized as 0 (an invalid hash value). Let $N$ be an intersection node. We compute its hash value $N.h = H(N.a.h \mid N.b.h)$. That is, the hash value for $N$ is the one-way hash value of the concatenation of the hash values stored in its child nodes.

To compute the hash values for all intersection nodes, we traverse the entire tree. First, create an empty stack $S$ and push the root of the tree to $S$. Then repeat the following operations until $S$ becomes empty:

* Peek a node $N$ from $S$;
* If the hash values in both nodes linked by $N.a$ and $N.b$ are valid, then
  - Set $N.h = H(N.a.h \mid N.b.h)$,
  - Pop the stack to remove $N$ from $S$.
* Otherwise,
  - If the hash value in the node linked by $N.a$ is invalid, push the node to the stack.
  - If the hash value in the node linked by $N.b$ is invalid, push the node to the stack.

***Step 4: Signing the tree***

As the last step of creating an IFMH-tree, we propose two approaches. The first approach is to sign only the root node, i.e., encrypting its hash value with the data owner's private key. We refer to this scheme as *one-signature*, alluding to the fact that the whole tree has only one signature.

The second approach, called *multi-signature*, is to create a signature for every subdomain node. Recall every subdomain node represented a subdomain that is determined by a set of inequality functions. For example, the set of inequality functions that determines the subdomain $S_4$ in Fig. 3 consists of $f_1(X) - f_2(X) < 0$, $f_1(X) - f_3(X) < 0$, $f_2(X) - f_3(X) \geq 0$, $f_1(X) - f_4(X) < 0$, $f_2(X) - f_4(X) \geq 0$, and $f_3(X) - f_4(X) < 0$. We do a one-way hash on these inequality functions, concatenate the result with the hash value stored the subdomain node, then do another one-way hash on the result. The final result is then signed with the data owner's private key.

## 3.2 Constructing Verification Objects

We now proceed to discuss how to process and construct verification objects for top-k, range, and KNN queries. While our discussion is limited to these three types, the IMH-tree can be used for other types of analytic queries. To process a query with function input $X$, the server first searches the IMH-tree to find the subdomain that contains $X$ and then searches the corresponding sorted function list for the functions that satisfy the query condition. Accordingly, the verification object built for a query has two objects, one for *subdomain verification* and the other for *function verification*.

Given the input $X$, the server searches the IMH-tree for the subdomain that contains $X$ and constructs the subdomain verification object at the same time. The algorithm is as follows.

* Let $N$ the tree root and $Q$ be an empty queue.
* As long as $N$ is not a subdomain node, repeat the following process:
  - Enqueue $N$ to $Q$.
  - Let $I_{p,q}$ be the intersection stored in $N$.
  - If $f_p(X) - f_q(X) \leq 0$,
    ○ Enqueue the node linked by $N.b$ to $Q$;
    ○ Set $N = N.a$;
  - Otherwise,
    ○ Enqueue the node linked by $N.a$ to $Q$;
    ○ Set $N = N.b$;

Once $N$ becomes a subdomain node, $Q$ will have all nodes in the search path and also their sibling nodes. Suppose $S_4$ in Fig. 4 is the subdomain that contains $X$. Then at the end of the search, $Q$ contains the nodes on the hash path $S_4$ along with their siblings. With the hash values stored in these nodes, one can recompute the hash value stored in the root node, and then use the data owner's public key to verify if these nodes are the part of the original tree where the search goes through. As such, in the case of the one-signature approach, the nodes in $Q$ form the subdomain verification object. In the case of the multi-signature approach, then the subdomain verification object is simply the set of inequality functions that determines the subdomain and the signature stored at the subdomain node.

Regardless the query being a KNN, range, or top-k query, the query result is a consecutive sub-list from the sorted function list linked by the subdomain node. Without loss of generality, let this sub-list be $f_a(X) \leq f_{a+1}(X) \leq \ldots \leq f_{b-1}(X) \leq f_b(X)$. The server locates $f_{a-1}$ and $f_{b+1}$, the two records that are immediate left and right to the sub-list, and constructs $VO(q)$ as follows. Let $N$ be the FMH node corresponding to $f_{a-1}$. It firstly adds $N$ and its sibling node to $VO(q)$, then set $N = N.p$ (i.e., move to $N$'s parent node). This process is repeated until $N$ becomes the root of the FMH node. In the end, all nodes along the hash path of the node corresponding to $f_{min}$ and their sibling nodes are added to $VO(q)$. The same process is used to add all nodes along the hash path of the node corresponding to $f_{b+1}$ and their corresponding sibling nodes to $VO(q)$. The data in $R(q)$ and $VO(q)$ are then transmitted to the user.

### 3.3 Verifying Query Results

A user submitted a query $q$ expects to receive a query result $R(q)$ and a verification object $VO(q)$, which includes an intersection verification object $IV(q)$, a function verification object $FV(q)$, and a digital signature $D$ that is created with the data owner's private key. If one-signature approach is used, this signature is the signed IFMH-tree root. Otherwise, it is the signed root of the FMH-tree linked by the subdomain that contains the query's function input $X$ is included.

The verification process is similar for both one-signature and multi-signature approaches. For the sake of brevity, we discuss only one-signature approach. The verification process consists of two steps. The first step is to verify the authenticity of the tree parts:

- Use the functions included $R(q)$ and nodes included $FV(q)$ to reconstruct the part of the FMH-tree.
- Use the nodes included in $IV(q)$ to reconstruct the part of the IMH-tree
- Recompute the hash value for the root node and encrypted it with the data owner's public key. Let this value be $M$.
- Decrypt the included signature with the data owner's public key. Let this value be $M'$.
- If $M$ and $M'$ are equal, then both parts of the IMH tree and the FMH tree are authentic. Otherwise, they are not, i.e., at least one node is forged, or at least one node in the original tree is not included.

If the tree parts are authentic, then the query issuer proceeds to the second step of verification, which is to mimic the server in query processing by searching the constructed tree parts. If the query result and the functions that are immediate left and right to the query result match those received from the server, the user can be assured that the functions included in the query result from the server are sound and complete.

## 4 PERFORMANCE STUDY

In this section, we study the performance of the proposed technique through security analysis, overhead analysis, and simulation.

### 4.1 Security Analysis

A user submitting a query $q$ expects to receive a query result $R(q)$, and a verification object that consists of $IV(q)$ for intersection verification and $FV(q)$ for function verification. With these data, the user recomputes the one-way hash value for the root node (either the root of IFMH tree in the case of one-signature approach or a subdomain node in the case of multi-signature approach). If the hash value matches the signature created by the data owner, the user can be assured that all data included in $R(q)$, $IV(q)$, and $FV(q)$ must be from the original IFMH-tree that is created by the data owner. This is guaranteed by the properties of the one-way hash and the way the root hash value is computed in both approaches. If any record or a hash value included is forged, the root hash value computed by the user would be different from the one computed by the data owner. As such, by checking if the subdomain included in the verification object contains the query's function input, the user can be assured if the records included in $R(q)$ are from the correct sorted list.

The question now is completeness. Let $R(q) = \{f_a, f_{a+1}, \ldots, f_b\}$ be the query result, $f_{a-1}$ and $f_{b+1}$ be respectively the immediate left and right records included in the verification object, and they do not satisfy the query condition, such that $f_{a-1}(X) < f_a(X) \leq f_{a+1}(X) \leq \ldots \leq f_b(X) < f_{b+1}(X)$ under the query's function input $X$. There are several cases $R(q)$ is not complete:

*Case 1:* Two records consecutive in $R(q)$ are not consecutive in the original sorted list. In other words, there exist two consecutive records $f_i$ and $f_{i+1}$ in $R(q)$ and at least one record $f_p$ in the original sorted list such that $f_i(X) \leq f_p(X) \leq f_{i+1}(X)$. Since the hash value of $f_p$ is part of the computation for the signature created by the data owner, the adversary would have to modify the functions in $R(q)$ for the user to recompute a hash value that matches. This is computation infeasible without knowing the data owner's private key.

*Case 2:* At least one boundary record is forged. Suppose a fake record $f'_{a-1}$ is created such that $f'_{a-1}(X)$ does not satisfy the query condition. This fake record is then inserted somewhere between $f_a$ and $f_b$, and so the query result becomes $\{f_i, f_{i+1}, \ldots, f_b\}$, i.e., functions from $f_a$ to $f_{i-1}$ are removed from $R(q)$. Since the forged record $f'_{a-1}(X)$ will be used by the user to compute the root hash value, the adversary must change some records in $R(q)$ to make this correct. Again, this is computation infeasible without knowing the private key used to create the signature.

### 4.2 Overhead Analysis

We analyze the overhead for data owner, server, and query issuers, respectively. We limit our discussion to linear functions, which are commonly used in analytic queries.

*Data owner overhead:* We first consider the size of an IFMH-tree, which determines the cost for uploading to the server. A set of $n$ linear functions with $d$ variables may create up to $O(n^2)$ intersections, which may partition a given domain into $O(n^{2d})$ subdomains. The IMH-tree is essentially a binary tree, so the number of total nodes for this tree is $O(n^{2d})$. For each subdomain, there is an FMH-tree, and since each FMH-tree is built on top of a list of $n$ sorted

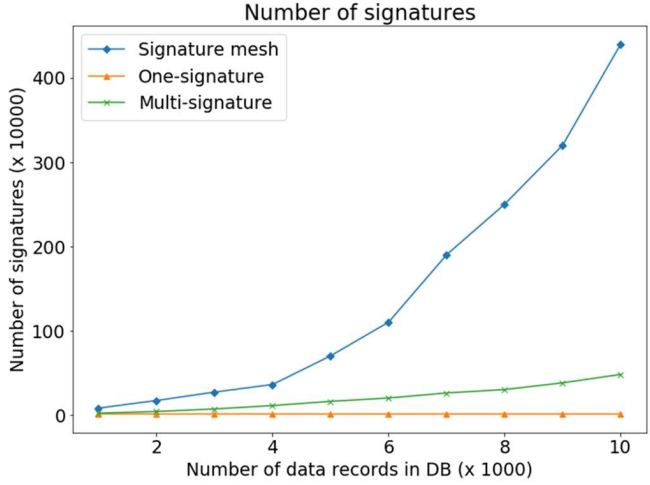

(a) Number of signatures needed for creating IFMH-tree/signature mesh

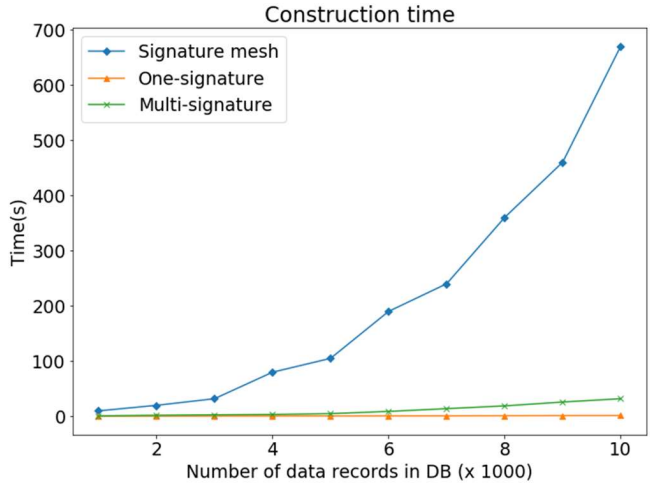

(b) Time of constructing the IFMH-tree/signature mesh

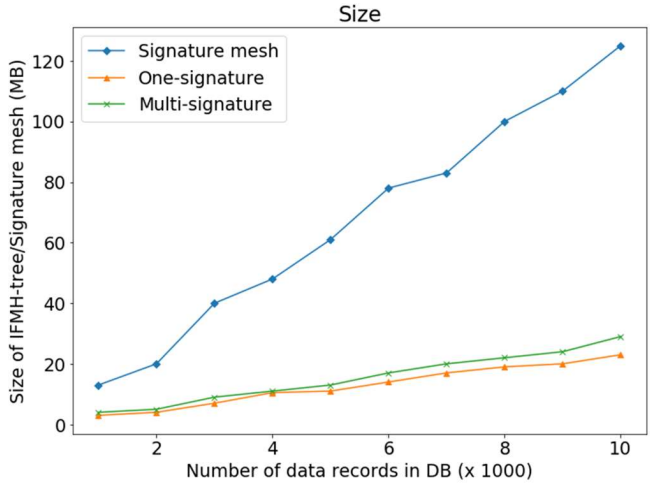

(c) Size of the IFMH-tree/signature mesh

Fig. 5. Data owner overhead

functions, the total number of nodes it has is $O(n)$. As such, the total nodes for a whole IFMH-tree is $O(n^{2d} + n^{2d} \times n) = O((n+1) \times n^{2d})$.

Now we analyze the cost of building an IMH-tree. Inserting an intersection may have to check against every node in the tree. So, the total number of nodes needed to be checked for inserting $O(n^2)$ intersections is $O(n^{2d})$. For each subdomain, the functions need to be sorted according to their scores. So, the total cost of sorting is $O(\,n \times \log n \times n^{2d})$. For each sorting list, building an FMH-tree requires to compute $O(n^2)$ times of one-way hash. So, the total hashes for the whole set of FMH-trees are $O(n^2 \times n^{2d})$. The total number of hashes for the IMH-tree is $O(n^{2d})$.

Building the IFMH-tree is still computation-intensive, but this is a one-time cost, and when compared with the original signature mesh, this cost has a significant reduction, given the fact that the number of signatures that need to be computed is much smaller.

*Server overhead:* The main advantage of our approach is in run time. Given a query with a function input $X$, the complexity of searching for the subdomain which contains $X$ is $O(\log n^{2d}) = O(d \log n)$, given the binary form of the IMH-tree. Since the $n$ (i.e., the number of records) is much larger than $d$ (i.e., the number of variables in ranking functions), this cost can be approximated as $O(\log n)$. In contrast, such a search in the signature mesh is linear. Once the subdomain is located, the server retrieves the query results in the corresponding sorted function list, which is also done in a binary search. So, the cost of finding the first function that satisfies the query condition is $O(\log n)$. Let $|q|$ be the query size, i.e., the number of functions in the query result. Then the total cost for query processing is $O(\log n) + |q|$. Since the verification object is constructed by way of the query processing, the whole cost is $O(\log n + |q|)$.

*User overhead:* Given a query $q$, the verification cost for the query issuer has to do the query size $|q|$. In both approaches, there is one signature included for a verification object. For multi-signature approach, the number of nodes in the FMH-tree is $O(|q|)$. So, the number of hashing that should be performed is $O(|q|)$. The one-signature approach, includes the IMH-tree part, which the cost is determined by the tree depth. In the worst case, this depth is $O(n^2)$ for a database of $n$ records. So, the total number of hashing in the verification is bounded to $O(n^2 + |q|)$.

## 4.3 Simulation Results

In this part, we study the performance of the proposed method by comparing it to signature mesh, in terms of the costs for data owner, server, and users. In most of our experiments, the number of records is considered 1,000 to 10,000 for two reasons. First, the cost of constructing a signature mesh increases exponentially with respect to the number of records, making it extremely time-consuming with our limited computing resources. Second, the limited data sizes are still clearly indicative of performance trends. Besides, this study is limited to linear ranking function, since it is the most common type used in analytic queries. The experiments were performed on a PC with 16GB of memory and an Intel i7 quad-core 3.1Gh processor.

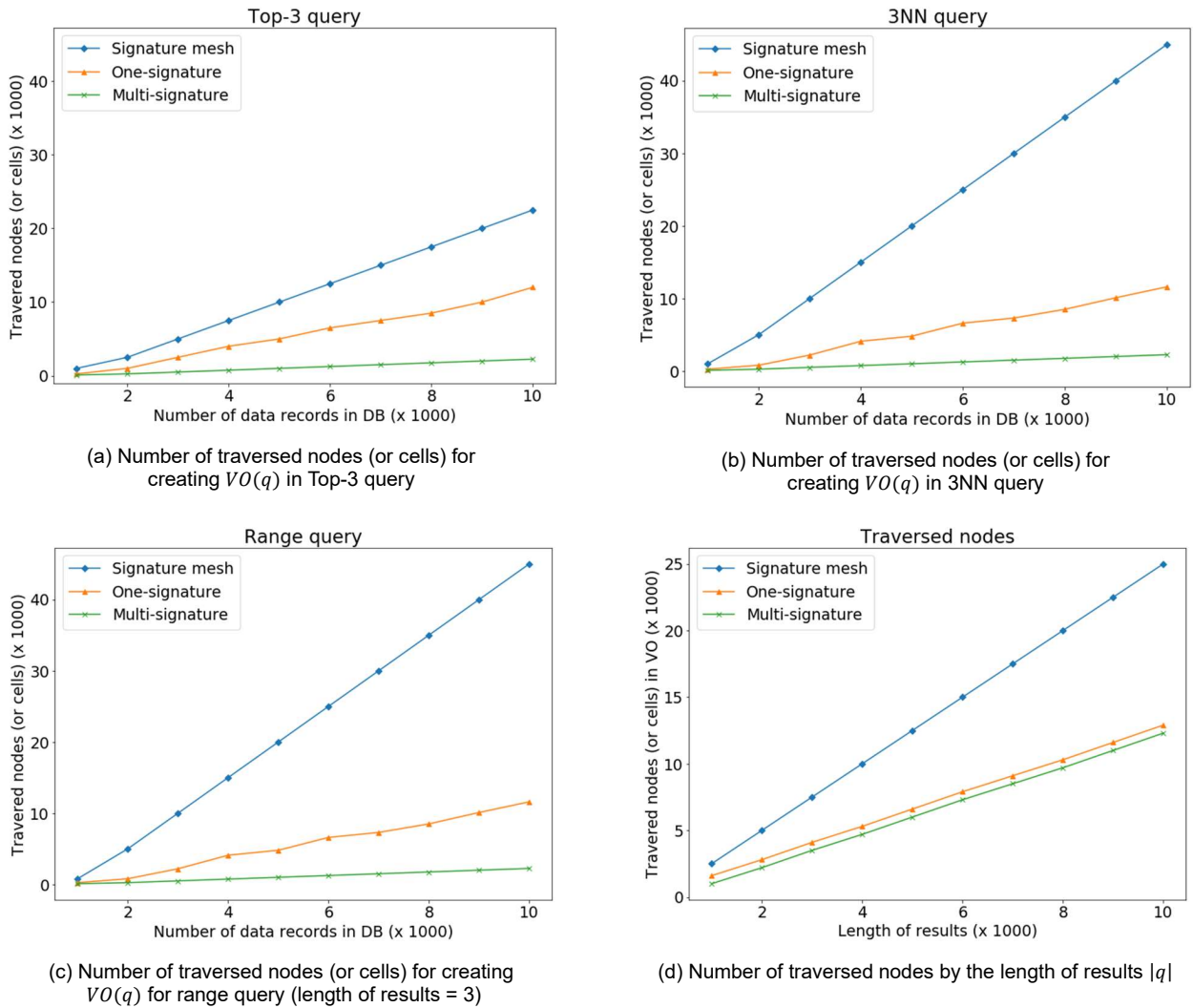

(a) Number of traversed nodes (or cells) for creating $VO(q)$ in Top-3 query

(b) Number of traversed nodes (or cells) for creating $VO(q)$ in 3NN query

(c) Number of traversed nodes (or cells) for creating $VO(q)$ for range query (length of results = 3)

(d) Number of traversed nodes by the length of results $|q|$

Fig. 6. Server overhead

### 4.3.1 Data owner overhead

The cost of the data owner consists of the computation time needed to construct the verification data structures – the IFMH-tree, and signature mesh. In signature mesh, the number of signatures can grow up to the number of sub-domains times the total number of records. In contrast, our one-signature approach has only one signature, while the multi-signature approach, this is equal to the number of the subdomains – Fig. 5a shows the number of signatures that each approach needs. The high number of signatures in the signature mesh also results in high computation time of creating the mesh (Fig. 5b), and its large size (Fig. 5c). The signature mesh needs more signatures, but has a smaller number of hash values when compared to our approaches. In this study, we use RSA for signatures, the size of which is 640 bytes, and SHA-256 for one-way hashing.

### 4.3.2 Server overhead

The server cost is measured by the number of the nodes in IFMH-tree or the number of the cells in the signature mesh that the server needs to traverse to process a query and construct a corresponding $VO(q)$. Fig. 6a-6c show the average cost for the server to process top-3, 3NN, and range queries whose results consists of only three records. As the number of records increases, the average cost under all approaches increases. This is not surprising since more data needs to be processed. In all settings, the signature mesh is the worst performer, and the performance gap increases as the number of records increases. This is due to the fact that this scheme performs a linear search in query processing and verification object construction. The performance of the two proposed approaches is quite stable. The one-signature approach incurs a relatively higher cost than the multi-signature approach. To build a verification object, the former requires to search both IMH-tree and FMH-tree, whereas the latter searches only the FMH-tree linked by the subdomain that contains the ranking function, which is much smaller.

In a separate study, the impact of the length of query results on the server cost is studied. For this purpose, the number of the records is set to 10,000; and the size of query results is considered 1,000 to 10,000. The performance results under the three approaches are plotted in Fig. 6d. Again, signature mesh always incurs a higher cost, while the performance of one-signature and multi-signature is quite comparable.

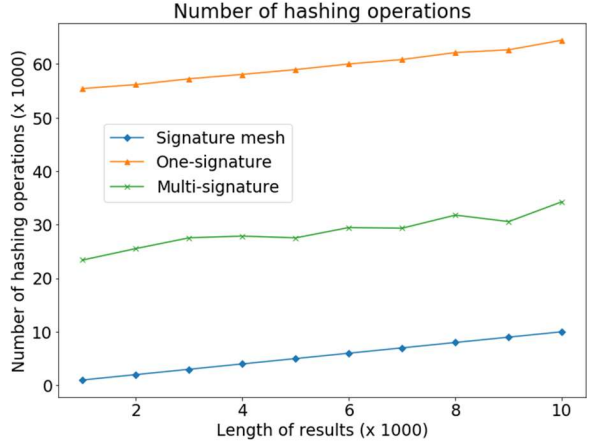

(a) Number of hashing operations for different lengths of query results

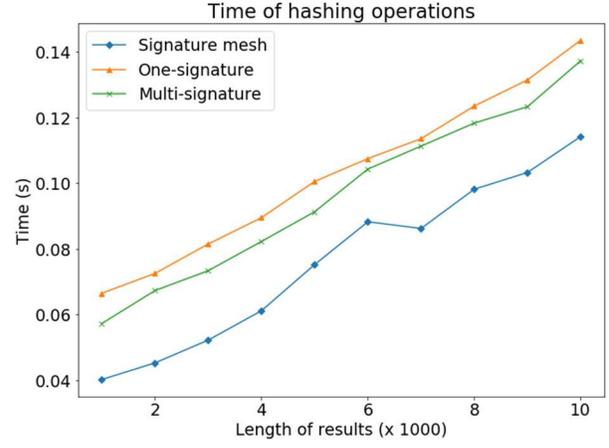

(b) Time of hashing operations for different lengths of query results

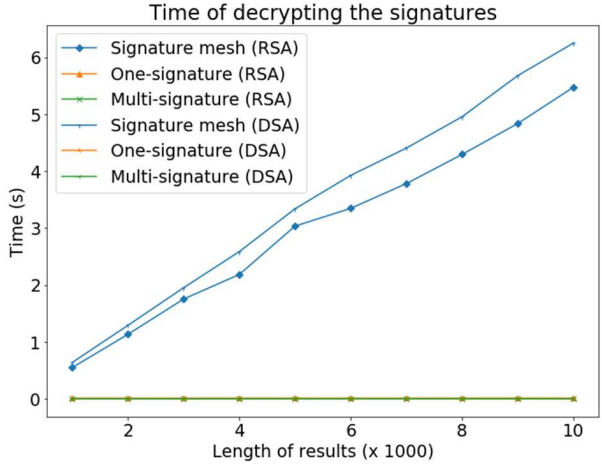

(c) Time of decrypting the signatures (RSA and DSA algorithms)

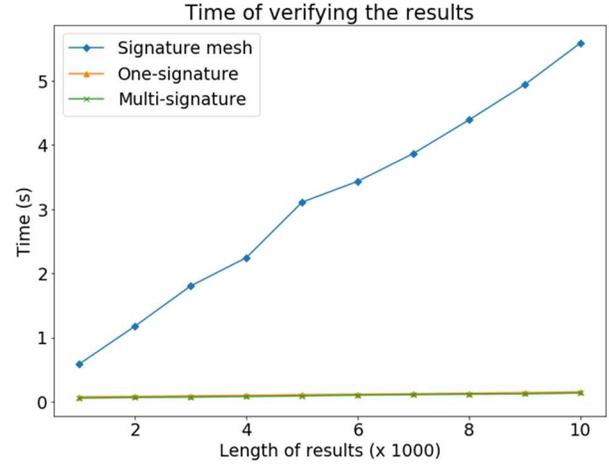

(d) Total time of verifying the results of the queries by the user

Fig. 7. User overhead for verifying the results of the queries

### 4.3.3 User overhead

In this experiment, the impact of the length of query result on the computations in users' side. The overall performance metric is considered to be the average time used in verifying query results. To provide a better understanding of this cost, we also collect other types of performance data, including the number of hashing operations, the actual time of performing hashing, and the time of decrypting signatures. The size of the query results varies from 1,000 to 10,000. The results are delineated in Fig. 7. As the length of the query result increases, the cost under all three approaches increases as well (Fig. 7a). Signature mesh has the best performance in terms of the number of hashing functions and the corresponding time (Fig. 7b). However, it has the most significant number of signatures, which incurs drastically more computation time in decrypting than one-way hashing. As such, the total time in verifying query results under signature mesh is a lot worse than the proposed approaches (Fig. 7d). These experiments also show that the performance gap increases as the query result size increases. Note that we use two different signature algorithms, RSA and DSA. Our results demonstrate the RSA is faster than DSA, but do not make a significant performance difference (Fig. 7c).

As for the two proposed approaches, multi-signature has a smaller number of hashing to perform than one-signature, while both have only one signature to be included in a query result. But the proposed approaches are not much different in terms of the total time incurred in verifying query results. This means that the time for hashing computation is tiny comparing to decrypting.

### 4.3.4 Communication overhead

The communication cost is determined by the size of a query result $|q|$ and the size of the corresponding verification object. Since the $|q|$ is the same for all three approaches, the only concern is the size of the $VO(q)$, which is studied through two experiments. First, the size of the database is fixed to 10,000 records, while the lengths of a query results vary from 1,000 to 10,000. Fig. 8a illustrates the difference of the sizes of verification objects. For

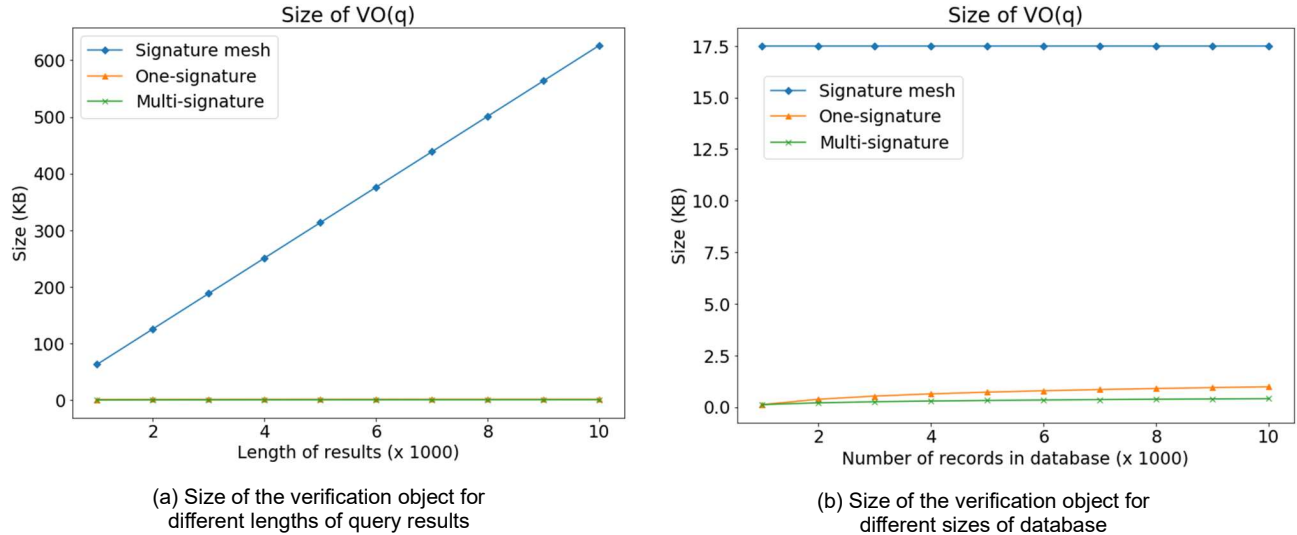

(a) Size of the verification object for
different lengths of query results

(b) Size of the verification object for
different sizes of database

Fig. 8. Communication overhead in terms of the size of $VO(q)$

signature mesh, there is a linear relationship between the length of the results and the size of the query. In contrast, this relationship is logarithmic for the two proposed techniques. The average sizes of $VO(q)$ for one- and multi-signature approaches are around 0.5KB and 1.3KB, respectively. In the second experiment, the length of a query result is fixed to 100, while the number of records varies from 1,000 to 10,000 (Fig. 8b). Increasing the number of the records does not affect the size of the $VO(q)$ in the signature mesh. As such, its curve is flat. In proposed approaches, a larger database results in a larger tree and, thus, a larger verification object, and the results confirm this fact. The increase, however, is slow for both approaches. The average size of $VO(q)$ in one-signature is larger than multi-signature. A verification object in one-signature approach needs to include some part of the IMH-tree, the height of which grows linear (in the worst case) with respect to the database size. Hence, the $VO(q)$ size is larger in one-signature approach; still the difference is not significant.

## 5 CONCLUSIONS

We have presented a generic data structure for verifying the correctness of the results of analytic queries. It has two components, namely IMH-tree and FMH-tree. The former is an extension of the existing I-tree, whereas, the latter, an extension of the existing MH-tree. We showed that the two components together support efficient execution and construction of verification objects of three representative types of analytic queries, namely top-k, range, and KNN queries. We proposed two versions of implementation, one-signature and multi-signature. The former signs the root for the IMH-tree, thus it has only one signature. In contrast, the latter creates a signature for every FMH-tree. The two approaches have their own pros and cons. However, when compared to the existing signature mesh, both approaches have significant performance improvements in terms of the overheads of data owner, server, and data user.

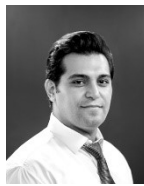

**Masoud Nosrati** received his B.S. and M.S. in 2012 and 2016, respectively, in computer engineering from Islamic Azad University, Iran. He is now a PhD student in the Department of Computer Science at Iowa State University. His current research interests include query processing and security in outsourced databases.

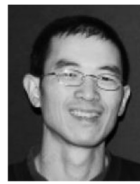

**Ying Cai** received his B.S. and M.S. degrees from Xi'an Jiaotong University, China. He graduated from the University of Central Florida with a PhD in computer science in 2002. Since 2003, Dr. Cai has been with the Department of Computer Science at Iowa State University. His current research interests include cloud computing, and privacy and security.